\documentclass[12pt]{article}
\usepackage{epsfig,amsfonts,amssymb}
\usepackage{hyperref}
\pdfoutput=1
\usepackage{cite}
\usepackage{cite}

\usepackage{comment}

\def\ZZZ{{\hbox{ Z\kern-1.6mm Z}}}
\def\RRR{{\hbox{ R\kern-2.4mm R}}}
\def\CCC{{\hbox{ C\kern-2.0mm C}}}
\def\zzz{{\hbox{z\kern-1mm z}}}

\newcommand{\qeq}{{\hbox{=\kern-2.3mm ? \kern.5mm }}}
\renewcommand{\qeq}{=}

\newcommand{\eps}{\epsilon}

\newcommand{\ve}{\varepsilon}

\newcommand{\II}{{\cal I}}

\newcommand{\OO}{{\cal O}}

\newcommand{\LL}{{\cal L}}

\newcommand{\wh}{\widehat}

\newcommand{\RR}{{\cal R}}
\newcommand{\NN}{{\cal N}}
\newcommand{\TT}{{\cal T}}

\newcommand{\be}{\begin{equation}}
\newcommand{\ee}{\end{equation}}
\newcommand{\ben}{\begin{eqnarray}\displaystyle}
\newcommand{\een}{\end{eqnarray}}

\newcommand{\refb}[1]{(\ref{#1})}
\newcommand{\p}{\partial}
\newcommand{\sectiono}[1]{\section{#1}\setcounter{equation}{0}}

\def\one{{\hbox{ 1\kern-.8mm l}}}
\def\zero{{\hbox{ 0\kern-1.5mm 0}}}

\newcommand{\bea}[1]{\begin{eqnarray}\label{#1} }
\newcommand{\eea}{\end{eqnarray}}

\newcommand{\eqref}{\refb}

\usepackage{bm}
\usepackage[table]{xcolor}

\newcommand{\Y}{Y}
\newcommand{\U}{\Upsilon}

\newcommand{\hh}{\hat h}

\begin{document}

\baselineskip 18pt

\thispagestyle{empty}
\begin{center}
    ~\vspace{5mm}
    
   \baselineskip=24pt
    
    {\Large \bf

 Supersymmetric Index for Half BPS Black Holes in N=2
 Supergravity with Higher Curvature Corrections
    }
    
    \vspace{0.4in}
    
    \centerline{\bf 
    Subramanya Hegde,$^1$  Ashoke Sen,$^2$
    P Shanmugapriya,$^{2,3}$ 
    Amitabh Virmani$^{4}$
    }

    \vspace{0.4in}

    $^1$ Max-Planck-Institut f\"ur Physik, Werner-Heisenberg-Institut, Boltzmannstr. 8, 85748 Garching bei M\"unchen, Germany
     \vskip1ex
    $^2$ International Centre for Theoretical Sciences, 
    Bengaluru - 560089, India \vskip1ex
    $^3$ Department of Physics, Indian Institute of Technology Madras, Chennai - 
    600036, India \vskip1ex
    $^4$ Chennai Mathematical Institute, 
    Kelambakkam, Tamil Nadu, India 603103    \vspace{0.1in}
    
    {\tt subbu@mpp.mpg.de,
    ashoke.sen@icts.res.in, shanmugapriya.prakasam@icts.res.in, avirmani@cmi.ac.in}
\end{center}

\vspace{0.4in}

\begin{abstract}

We compute the supersymmetric index of half BPS black holes 
in N=2 supergravity with higher curvature corrections and show that the result
agrees with the degeneracy of supersymmetric extremal black holes carrying the
same charges. Both sides of the computation are done gravitationally.

\end{abstract}

\pagebreak

\baselineskip=18pt

\tableofcontents

\sectiono{Introduction and summary}

Historically, testing Bekenstein-Hawking formula and its generalization in string theory
involved
comparison of black hole entropy, computed via the Bekenstein-Hawking-Wald
formula, possibly modified by quantum corrections, with an appropriate
supersymmetric index of microscopic
states in string theory carrying the same set of charges as the black 
hole\cite{9601029,9711053,9812082,1008.3801}. 
This relied on the assumption that for the supersymmetric black holes the index
and the degeneracy are the same.
Recently a method has been developed for computing the index directly on the
black hole side\cite{2107.09062} (see \cite{1810.11442} for earlier work on $AdS$
black holes and \cite{2011.01953} for related work). 
So far the result of this new computation has agreed with the entropy
computed via Bekenstein-Hawking formula, together with logarithmic corrections
computed from quantum corrections, confirming the conjectured agreement between 
the index and entropy for macroscopic black 
holes\cite{2306.07322,2308.00038,2310.07763,2311.09427,2401.13730,2402.03297}. 
Our goal in this paper will be to
extend this test to theories with higher derivative terms.

We shall be working with the classical $\NN=2$ supergravity
action, containing higher derivative terms, in asymptotically flat
four space-time dimensions\cite{deRoo:1980mm,deWit:1980lyi,
Bergshoeff:1980is,deWit:1982na,deWit:1996gjy}. 
In this case the prescription for computing the index can be stated as 
follows\cite{2107.09062}.
We construct a rotating black hole solution, carrying the same charges for which we
want to compute the index, with the rotation parameter adjusted so that the
angular velocity $\Omega_H$ at the horizon is related to the inverse temperature
$\beta$ via the relation
\be\label{eombeta}
\Omega_H\, \beta=2\pi i\, .
\ee
Let $S_{wald}$ be the Wald entropy\cite{9307038,9312023} of such a black hole. 
The chemical potential $\Omega$ dual to the angular momentum $J$, defined
as
\be
\Omega= \frac{1}{\beta} {\p S_{wald} \over \p J}\, ,
\ee
takes the value
\be
\Omega = -\Omega_H= -2\pi i/\beta\, .
\ee
Then the log of the index is
given by,
\be\label{eindex}
I = S_{wald} + 2\pi i\, J\, .
\ee 
Our goal in this paper
will be to compute \refb{eindex} explicitly
and show that the result agrees with the entropy of the 
spherically symmetric
BPS black hole solution carrying the same set of charges\cite{9812082}. This 
tests the
equality of the index and degeneracy.

For our analysis we shall make heavy use of the black hole solution found in 
\cite{2310.07763,2402.03297}, based on the earlier papers \cite{9705169,0009234}.
In particular the black hole solution relevant for our analysis was given in 
\cite{2402.03297}. The main reason that \cite{2402.03297} was not able to compute
\refb{eindex} and compare with the entropy of extremal black holes is that the
solution was not known explicitly, but given in terms of a function that satisfies a
second order partial differential equation. Our main observation is that the Wald
entropy is given by an integral of a total derivative term at the horizon and can be
expressed in terms of the boundary contribution from the poles. The values of the
various fields at the poles are known explicitly from the results of 
\cite{0009234,2310.07763,2402.03297}. This allows us to compute \refb{eindex}. Writing the Wald entropy as a total derivative is similar in spirit to 
the equivariant
localization in supergravity discussed in \cite{2306.03868}.

\sectiono{Review of known results} \label{s2}

In this section we shall collect known results that will be used in our analysis.
Our main source will be \cite{0009234,2310.07763,2402.03297, Mohaupt:2000mj}. 
We note at the outset that \cite{0009234, Mohaupt:2000mj}
uses a sign convention for the Riemann tensor that is opposite of the standard one,
and as a result the Einstein-Hilbert action has a minus sign multiplying 
$\int d^4x \sqrt{-\det g} R$. We shall use the same convention in our analysis.

We shall find the formula for the Wald entropy by working with the 
Lorentzian metric even though the Lorentzian solution is complex.
We shall 
use the convention that $\mu,\nu,\rho,\cdots$ are four dimensional curved
indices and
$a,b,c,\cdots$ 
are four dimensional tangent space indices. 
We use mostly plus signature metric and 
our convention for the epsilon tensors in the tangent space will be,
\be
\ve^{0123}=1, \qquad \ve_{0123}=-1\, .
\ee
The relevant fields that will enter our analysis include a set of complex 
scalars $\{X^I\}$, an auxiliary
real scalar field $D$, an anti-self-dual, complex, anti-symmetric tensor field $T_{ab}$
and its complex conjugate self-dual tensor field $\bar T_{ab}$,\footnote{In the
notation of \cite{0009234, Mohaupt:2000mj} $T_{ab}$ was called $T^{ij}_{ab}\varepsilon_{ij}$ and 
$\bar T_{ab}$ was called $T_{abij}\varepsilon^{ij}$.}
a composite real scalar field $\chi$ constructed from other elementary scalars
and a set of gauge fields $A_\mu^I$.
There are other fields in the theory that will not be needed for our analysis.
$T_{ab}$ and $\bar T_{ab}$ satisfy the (anti-)self-duality condition:
\be\label{etadef}
T^{ab} = -{i\over 2}\, \ve^{abcd} T_{cd}, \qquad \bar T^{ab} = 
{i\over 2} \, \ve^{abcd} \bar T_{cd}\, ,
\ee
Let $F^I_{\mu\nu}=\p_\mu A^I_\nu - \p_\nu A^I_\mu$ be the  field strength
associated with the gauge field. We define:
\be\label{ehodge}
F^{-I}_{ab} = {1\over 2} \left( F^{I}_{ab} -{i\over 2} \ve_{abcd} F^{Icd}\right), 
\ee
and
\be
{\hat A} = T_{ab} T^{ab}.
\ee
The form of the Lagrangian is controlled by a function $F(\{X^I\}, {\hat A})$, known as
the prepotential, which
satisfies the homogeneity relation,
\be\label{ehomx}
F(\{\lambda X^I \}, \lambda^2 {\hat A}) = \lambda^2 \, F(\{X^I\}, {\hat A})\, .
\ee
We shall define:
\be
F_I ={\p F\over \p X^I}, \qquad F_{\hat A} ={\p F\over \p {\hat A}}\, .
\ee

We shall now write down the action for $\NN=2$ supergravity in four
dimensions. For reasons to be explained below, 
we shall only be interested in the terms that depend on the Riemann
tensor and / or the auxiliary $D$ field of supergravity.  For this reason we shall 
denote the terms that do not depend
on $D$ or the Riemann tensor by $\cdots$ and will not explicitly write down
their form.  The bosonic part of the action takes the form,
\be\label{eactionpre}
S = \int d^4x\, \sqrt{-\det g} \, \LL\, ,
\ee
where\cite{deRoo:1980mm,deWit:1980lyi,
Bergshoeff:1980is,deWit:1982na,deWit:1996gjy},
\ben\label{elag}
8\pi\LL &=& \bigg[-i F_I \bar X^I \left({1\over 6} R - D\right) 
+{1\over 2} \, i\, \wh F^{-ab} \, F_{{\hat A}I} \, \left(F^{-I}_{ab} -{1\over 4}
\bar X^I \, T_{ab}\right) +{1\over 2} i \, F_{\hat A}\, \wh C 
+{i\over 4}F_{{\hat A}{\hat A}} \wh F^{-ab} \, \wh F^{-}_{ab} \, 
\nonumber \\ &+& {\rm c.c.}\bigg] 
+ \chi\left({1\over 6}R +{1\over 2} D\right) + \cdots\, ,
\een
where bar denotes complex conjugate field, and, 
\be
\wh{C} = 64{\mathcal{R}(M)^-}_{ef}{}^{gh}\mathcal{R}(M)^{-ef}{}_{gh}+16 \, f_a{}^c\, 
T^{ab} \bar T_{cb}+\cdots,
\ee
\be
\wh{F}^{-ab} =-16\mathcal{R}(M)_{cd}{}^{ab}T^{cd},
\ee
\be
f_\mu{}^a  = {1\over 2} R_\mu{}^{a} - {1\over 4} \left(D+{1\over 3}R\right) e_\mu{}^a+\cdots,
\ee
\be \label{ecurlyRdef}
\mathcal{R}(M)_{ab}{}^{cd}=R_{ab}{}^{cd}-2R_{[a}^{[c}\delta^{d]}_{b]}+{R\over 3}\delta^c_{[a}\delta^d_{b]}
+D\delta^c_{[a}\delta^d_{b]}+ \cdots\, ,
\ee
\be
\mathcal{R}(M)^{-ef}{}_{gh} = {1\over 2}
\left(\mathcal{R}(M)^{ef}{}_{gh} -{i\over 2} \ve_{ghab}\mathcal{R}(M)^{efab}\right)
\, .
\ee
c.c. in \refb{elag} denotes complex conjugation that acts on the fields as follows.
In the description of the field content of Lorentzian supergravity 
given above, we have some real fields 
({\it e.g.} the metric and $D$) and some
complex fields ({\it e.g.} $X^I$ and $T^{ab}$). c.c. will keep the real fields unchanged
and take the complex fields to their complex conjugate fields ({\it e.g.} $\bar X^I$ and 
$\bar T^{ab}$). 
Any other $i$ that appears explicitly in \refb{etadef}-\refb{ecurlyRdef}
will be changed to $-i$ under c.c.
Later in our analysis we shall encounter complex solutions in Lorentzian
supergravity in which some of the real fields ({\it e.g.} the metric) take complex
values. c.c. {\it does not complex conjugate the solution, i.e.\ acting on
$g_{\mu\nu}$ it will give us back $g_{\mu\nu}$ even if $g_{\mu\nu}$ is
complex}.

We now make some comments on the missing terms, denoted by the $\cdots$
in \refb{elag}-\refb{ecurlyRdef}.
For computing the Wald entropy we need to find the derivative of the action with
respect to the Riemann tensor at fixed values of other fields. For this computation
we do not need the missing terms since they are independent of the Riemann
tensor. However the expression for the Wald entropy obtained this way will involve
$\RR(M)_{abcd}$ and we need the full expression of $\RR(M)_{abcd}$ for
computing the entropy. Fortunately the expressions for $\RR(M)_{abcd}$ 
for the solution that we shall
consider have been given in \cite{0009234} and we can simply use those results
without having to 
know the actual expression for $\RR(M)_{abcd}$ in terms of the
fundamental fields.

The action \refb{eactionpre},
\refb{elag} has a gauge symmetry that allows us to scale all the $X^I$'s
by an arbitrary function of the space-time coordinates. The other fields also transform
according to fixed rules. Following \cite{0009234} we shall work in the gauge
\be
\chi=-2\, ,
\ee
that eliminates the freedom of doing these scalings.

Ref.~\cite{0009234} constructed a class of stationary
solutions in this theory preserving
$\NN=1$ supersymmetry.
We shall now collect all the relevant results from \cite{0009234}.
The metric takes the form,
\be\label{emetric}
ds^2=-e^{2g}(dt+  \sigma_\alpha \, dx^\alpha)^2+e^{-2g} ds^2_{\mathrm{base}},
\ee
where
$ds^2_{\mathrm{base}}$
is the metric on the flat three dimensional base space 
and $g$ is a smooth function of the base space
coordinates.
The three dimensional vector field
$\sigma_\alpha$ also depends only on the base space coordinates.
We shall parametrize the base space by the coordinates $x^\alpha$,
and denote by
$\alpha,\beta,\gamma,,\cdots$ the three dimensional curved indices
associated with these coordinates. 
We shall also use the convention that
$m,n,p,\cdots$ are three dimensional tangent space indices, taking values 
1, 2 and 3 and choose the convention for the three dimensional epsilon
tensor $\ve^{mnp}$ such that,
\be
\ve^{123}=1\, .
\ee
We take the vielbeins as
(eq.(5.21) of \cite{0009234}):
\be
e_\mu^{~0}  dx^\mu= e^g (dt +  \sigma_\alpha \, dx^\alpha), \qquad 
e_\mu^{~p} dx^\mu
= e^{-g} \, \hat e_\alpha^{~p}\,  dx^\alpha \quad \hbox{for $p=1,2,3$}\, ,
\ee
where $\hat e_\alpha^{~p}$'s are the vielbeins associated with the three dimensional
base metric.
We define $\hat E_p^{~\alpha}$ to be the
inverses of $\hat e_\alpha^{~p}$'s.

We now introduce some further definitions.
We define
(eq.(5.27) of \cite{0009234}):
\be \label{e57}
\TT_\alpha \equiv {1\over 4} \, e^{-g}\, \hat e_\alpha^{~p} \, \bar h\, T_{p0}\, ,
\ee
where $h$ is a position dependent phase that appears in the 
construction of the solution.
We denote by $\nabla_\alpha$ the three dimensional covariant derivative
computed using the vielbein $\hat e_\alpha^{~p}$.
We also introduce the tangent space quantities,
\be \label{edeftm}
\hat\nabla_p \equiv \hat E_p^{~\alpha} \nabla_\alpha\, , \qquad
T_p \equiv \hat E_p^{~\alpha} \TT_\alpha\, .
\ee
Finally we
define (eq.(5.24) of \cite{0009234}):
\be\label{e226x}
R(\sigma)^m = \ve^{mnp} \hat E_n^{~\alpha} \hat E_p^{~\beta} \p_\alpha \sigma_\beta\, .
\ee
We can invert \refb{e57} as,
\be\label{e509}
\bar h\, T_{p0} = 4\, e^g\, T_p\, .
\ee
The anti-self-duality of
$T_{ab}$ then determines the components $\bar h\, T_{pq}$:
\be\label{e510}
\bar h \, T_{pq}= -i\, \ve_{pq0r} \, \bar h\, T^{0r} \
=4\, i\, e^g \, \ve_{pq}^{~~r}\, T_r\, .
\ee
We also have the complex conjugate relations:
\be\label{e511}
h\, \bar T_{p0} = 4\, e^g\, \bar T_p, \qquad h \bar T_{pq}
 =- 4\, i\, e^g \, \ve_{pq}^{~~r}\, \bar T_r\, .
\ee

It was shown in \cite{0009234}
(eq.(5.28))  that for the supersymmetric solution studied there, we have,
\be \label{e511x}
T_m =\hat\nabla_m g -{1\over 2}\, i\, K_m, \qquad K_m\equiv 
 e^{2g} \, R(\sigma)_m
= 
e^{2g} \, \ve_{m}^{~~np} \hat E_n^{~\alpha}  \hat E_p^{~\beta} \p_\alpha \sigma_\beta\, .
\ee
Its complex conjugate relation is,
\be \label{e511xx}
\bar T_m =\hat\nabla_m g + {1\over 2}\, i\, K_m\, .
\ee
Note that $K_m$ is not complex conjugated even though
for a complex metric of the kind we shall consider,
$K_m$ will be complex.
It was also shown in  \cite{0009234}  (eqs.(5.19) and (5.30)) that
for the supersymmetric solution, the components
of $\RR(M)$ are given by:\footnote{Note that these formulae include 
the contribution from the dots in \refb{ecurlyRdef}.}
\ben \label{e512}
\RR(M)_{pq0r} &=& {1\over 2} \, i\, \ve_{pq}^{~~s}\, 
e^{2g}\, \left[\hat\nabla_r T_s + 2 T_r T_s
-\delta_{rs} T_m T_m\right] +{\rm c.c.}\, ,
 \nonumber \\
\RR(M)_{0rpq} &=& {1\over 2} \, i\, \ve_{pq}^{~~s}\, 
e^{2g}\, \left[\hat\nabla_s T_r + 2 T_r T_s
-\delta_{rs} T_m T_m\right] +{\rm c.c.}\, , \nonumber \\
\RR(M)_{0p0q} &=& -{1\over 2}\, e^{2g}\, \left[\hat\nabla_q T_p + 2 T_q T_p
-\delta_{pq} T_m T_m\right] +{\rm c.c.} \, , \nonumber \\
\RR(M)_{pqrs} &=&  {1\over 2}\, e^{2g}\, \ve_{rs}^{~~v} \ve_{pq}^{~~u} 
\left[\hat\nabla_v T_u + 2 T_v T_u
-\delta_{uv} T_m T_m\right]+{\rm c.c.}
\, . 
\een
For computing the Wald entropy we also need an expression for $F^{-I}_{0p}$. 
We get from (5.33)
 of \cite{0009234},
\be \label{e135x}
F^{-I}_{0p} = -e^g \left[\hat\nabla_p(\bar h X^I)+(\hat\nabla_p g) \, h\bar X^I-{1\over 2}\, i\, e^{2g}\,
R(\sigma)_p \, (\bar h X^I + h\bar X^I)\right]\, .
\ee

So far we have not specified the solutions for $X^I$, $\sigma_m$ and $g$. 
It was shown in \cite{0009234} 
(eq.(6.5)) that we have,
\be\label{eharm}
-i\, e^{-g} (\bar h X^I - h\bar X^I) = H^I, \qquad
-i\, e^{-g} (\bar h F_I - h\bar F_I) =H_I\, ,
\ee
where $H^I$ and $H_I$ 
are harmonic functions on the base. Furthermore, $g$ and $\sigma_\alpha$ 
are determined
using the equations\cite{0009234} (eqs.(5.38), (5.39)):
\ben\label{e232x}
i\, (\bar X^I F_I- X^I\, \bar F_I) +{1\over 2}\chi &=& -128 \, i \,  e^{3g}\, \hat\nabla^p 
\left[e^{-g} \, \hat\nabla_p g\, (F_{\hat A}-\bar F_{\hat A})\right] - 32\, i\, e^{6g}\,
 (R(\sigma)_p)^2 (F_{\hat A}-\bar F_{\hat A})
\nonumber \\ 
&&\ - 64\, e^{4g}\, R(\sigma)_p \, \hat\nabla^p (F_{\hat A}+\bar F_{\hat A})\, ,
\een
and
\ben \label{e233x}
&& (\bar h \, X^I - h \, \bar X^I) \hat\nabla_p (\bar h \, F_I - h \, \bar F_I)
- (\bar h \, F_I - h \, \bar F_I) \hat\nabla_p (\bar h \, X^I - h \, \bar X^I)
-{1\over 2} \, \chi\, e^{2g}\, R(\sigma)_p \nonumber \\
&=& 128\, e^{2g}\, \hat\nabla^q \left[ 2 \hat\nabla_{[p}g \hat\nabla_{q]}(F_{\hat A}+\bar F_{\hat A})
+ i\, \hat\nabla_{[p} \left(e^{2g}R(\sigma)_{q]}(F_{\hat A}-\bar F_{\hat A})\right)\right]\, .
\een

We shall specialize to  axially symmetric solutions.
In this case it is natural to 
use the Boyer-Lindquist coordinates to describe the three dimensional base space
so that we have,
\be\label{embase}
ds^2_{\mathrm{base}}={{r^2-b^2\cos^2\theta} \over {r^2-b^2}}dr^2+(r^2-b^2\cos^2\theta)d\theta^2+(r^2-b^2)\sin^2\theta d\phi^2,
\ee
and
\ben
&& \hat e_\alpha^{~1}\,  dx^\alpha = dr 
\left( {r^2 - b^2\cos^2\theta\over r^2-b^2}\right)^{1/2},  \qquad
\hat e_\alpha^{~2}\,  dx^\alpha = (r^2 - b^2\cos^2\theta)^{1/2}\, d\theta, \nonumber \\
&&
\hat e_\alpha^{~3}\,  dx^\alpha = (r^2-b^2)^{1/2} \, \sin\theta d\phi\, .
\een
The relation between the $r,\theta,\phi$ coordinates and the usual Cartesian 
coordinates has been given in appendix \ref{sb}. Also, requiring
invariance under $(t,\phi)\to (-t,-\phi)$, we can parametrize $\sigma_\alpha$ as,
\be
\sigma_\alpha \, dx^\alpha=\omega_\phi \, d\phi\, .
\ee
The inverse four dimensional vielbeins are,
\ben\label{einvfour}
&&\hskip -.5in  E_0^\mu {\p\over \p x^\mu} =
e^{-g}\, {\p\over \p t}, \qquad 
E_1^{~\mu} {\p\over \p x^\mu} = e^g\, \left( {r^2 - b^2\cos^2\theta\over r^2-
b^2}\right)^{-1/2}{\p\over \p r}, \nonumber \\ && \hskip -.5in 
 E_2^{~\mu} {\p\over \p x^\mu} =  e^g\, (r^2 - b^2\cos^2\theta)^{-1/2} 
{\p\over \p \theta}, \qquad
 E_3^{~\mu} {\p\over \p x^\mu} = {e^g\over (r^2-b^2)^{1/2}
\sin\theta} \left( {\p\over \p \phi}-\omega_\phi\, {\p\over \p t}\right) .
\een
The inverse three dimensional vielbeins are,
\ben\label{e56x}
&& \hat E_1^{~\alpha} {\p\over \p x^\alpha} = \left( {r^2 - b^2\cos^2\theta\over r^2-
b^2}\right)^{-1/2}{\p\over \p r}, \qquad
\hat E_2^{~\alpha} {\p\over \p x^\alpha} = (r^2 - b^2\cos^2\theta)^{-1/2} 
{\p\over \p \theta}, \nonumber \\ &&
\hat E_3^{~\alpha} {\p\over \p x^\alpha} = {1\over (r^2-b^2)^{1/2}
\sin\theta} {\p\over \p \phi}\, .
\een

One can show that the finiteness of the Ricci scalar associated with the metric
\refb{emetric} requires that $\omega_\phi$ is constant on the $r=b$ surface\cite{2305.08910} (eq.(6.5)):\footnote{Note that $\omega$ of \cite{2305.08910}
is $-1/\omega_\phi$ in our notation.}
\be\label{e125x}
\p_\theta\omega_\phi = 0 \qquad \hbox{at $r=b$}\, .
\ee
In this geometry the horizon is situated at $r=b$ where the Killing vector,
\be\label{e57x}
{\p\over \p t} - \omega_\phi^{-1} {\p\over \p\phi}
\ee
has vanishing norm. This gives the angular velocity $\Omega_H$ at the 
horizon to be,
\be\label{eAng}
\Omega_H=-\omega_H^{-1}, \qquad \omega_H\equiv \omega_\phi|_{r=b}\, .
\ee
Comparison with \refb{eombeta} now tells us that $\omega_H$ must be
purely imaginary.

For solution that is
of interest to us, \refb{eharm}
takes the form\cite{2310.07763,2402.03297}:
\ben \label{e242}
-i\, e^{-g} (\bar h X^I - h\bar X^I) &=& \hh^I + {\gamma_N^I\over r_N} + 
{\gamma_S^I\over r_S} \, ,
\nonumber \\
-i\, e^{-g} (\bar h F_I - h\bar F_I) &=&\hh_I + {\gamma_{NI}\over r_N} + 
{\gamma_{SI}\over r_S} \, ,
\een
where $\hh^I$ and $\hh_I$ are constants and
\be\label{ea2y}
r_N=r-b\cos\theta, \qquad r_S=r+b\cos\theta\, ,
\ee
are the distances to the point $(r,\theta,\phi)$ from the north and the south poles
of the horizon,
measured in the base metric (see \refb{ea2x}).
$\gamma_N^I$, $\gamma_{NI}$, $\gamma_S^I$ and $\gamma_{SI}$ are determined
from the equations:
\be\label{egammasum}
\gamma_N^I +  \gamma_S^I = P^I, \qquad
\gamma_{NI} +  \gamma_{SI} = Q_I\, ,
\ee
where $P^I$ and $Q_I$ are the electric and the magnetic charge vectors carried by the
black hole, and demanding that,
\ben \label{e242new}
&&- i\, e^{-g} \bar h X^I  = {\gamma_N^I\over r_N} + \OO(1),\quad 
 - i\, e^{-g}\, \bar h F_I =  {\gamma_{NI}\over r_N}  +\OO(1), 
\quad e^{-2g}\, \bar h^2 \, {\hat A} = -{64\over r_N^2},
\qquad \hbox{as $r_N\to 0$}\, ,
\nonumber \\
&& i\, e^{-g}   h\bar X^I = 
\OO(1)\, , \quad 
 i\, e^{-g}  \, h\bar F_I =
\OO(1)\, , \quad e^{-2g}\,  h^2 \, \bar {\hat A} = \OO\left({1
\over r_N}\right),
\qquad \hbox{as $r_N\to 0$}\, ,\nonumber \\
&& i\, e^{-g}   h\bar X^I = 
{\gamma_S^I\over r_S} +\OO(1)\, , \quad 
 i\, e^{-g}  \, h\bar F_I =
{\gamma_{SI}\over r_S} +\OO(1)\, , \quad e^{-2g}\,  h^2 \, \bar {\hat A} = -{64
\over r_S^2},
\qquad \hbox{as $r_S\to 0$}\, ,\nonumber \\
&& - i\, e^{-g} \bar h X^I  = \OO(1),\quad 
 - i\, e^{-g}\, \bar h F_I = \OO(1), 
\quad e^{-2g}\, \bar h^2 \, {\hat A} = \OO\left({1\over r_S}\right),
\qquad \hbox{as $r_S\to 0$}\, . 
\een
Note that since we are allowing complex solutions, $\bar X^I$ is not the
complex conjugate of $X^I$.
Since for given $\{X^I\}$ and $\hat A$, $F_I$ is known, \refb{e242new} determines
$\gamma_{NI}$ in terms of $\{\gamma_N^I\}$ and  $\gamma_{SI}$ in terms of 
$\{\gamma_S^I\}$. These relations, together with \refb{egammasum}, determine
$\gamma_N^I$, $\gamma_{NI}$, $\gamma_S^I$ and $\gamma_{SI}$ completely.
In fact, it follows from these equations and the old attractor equations of
\cite{0009234} that 
the $i\gamma_N^I$'s give the
attractor values of the fields $X^I$ and
the $-i\gamma_S^I$'s give the attractor values of the fields $\bar X^I$
for spherically symmetric
extremal black holes if we work in a gauge in which $\hat A=-64$  
such that $\bar{h} \,  e^{-g} = \frac{1}{r_N}$ as $r_N \to 0$ and $h \,  
e^{-g} = \frac{1}{r_S}$ as $r_S \to 0$.

In order that the metric computed from \refb{e232x}, \refb{e233x}
does not have a Dirac string singularity, we also need to impose
the constraint\cite{2310.07763,2402.03297}:\footnote{This equation is obtained
by taking the difference between eq.~(4.23) of \cite{2402.03297} and its complex
conjugate that also exchanges $N$ and $S$. This
condition can also be obtained as follows: if we assume that $\omega_\phi$ 
falls off as $1/r$ for large $r$ so as to have
a metric satisfying regular asymptotic fall off condition, then the requirement that the coefficient of $1/r^2$ term of \refb{e233x} be zero gives \eqref{ehhgam}. Since from \refb{e242} we see that $\hh^I$ is related to the asymptotic
values of the scalars and since the asymptotic values of the physical scalars form
part of input data, it may seem surprising that we have a condition on $\hh$. However note that for complex solutions of the type we are
considering, specifying the asymptotic values of the physical
scalars leaves open the possibility of scaling the asymptotic values of the $X^I$'s by
an arbitrary complex constant and the asymptotic values of the $\bar X^I$'s by another
arbitrary complex constant. Using this freedom, we can ensure that 
\refb{ehhgam} holds. This fixes one combination of the two complex constants; the
other combination is fixed by demanding that the coefficient of the constant term
in the left hand side of \refb{e232x} vanishes for large $r$.}
\be\label{ehhgam}
 \langle \hh, \gamma_N+\gamma_S\rangle =0\, ,
\ee
where the angle bracket $\langle,\rangle$ between two
vectors $a$, $b$ is defined as
\be
\langle a, b\rangle \equiv a^I b_I - a_I b^I\, .
\ee
The absence of Dirac string singularity can also be used to show that the
temperature of the black hole satisfies the equation \refb{eombeta}.

It was also shown in \cite{2402.03297} that near the poles on the horizon the leading
contributions to \refb{e232x}, \refb{e233x} 
come from the left hand side. Hence we have,
\be \label{e232xy}
i\, (\bar X^I F_I- X^I\, \bar F_I) +{1\over 2}\chi \simeq 0\, ,
\ee
\be \label{e233xy}
(\bar h \, X^I - h \, \bar X^I) \hat\nabla_p (\bar h \, F_I - h \, \bar F_I)
- (\bar h \, F_I - h \, \bar F_I) \hat\nabla_p (\bar h \, X^I - h \, \bar X^I)
-{1\over 2} \, \chi\, e^{2g}\, R(\sigma)_p 
\simeq 0\, ,
\ee
near the poles at the horizon.

Finally, using the relation between $\gamma_N$, $\gamma_S$ and the old attractors
values of $X^I$, $F_I$, $\bar X^I$ and $\bar F_I$ in the $\hat A=-64$ gauge, 
it was shown in \cite{2402.03297} that the spherically symmetric extremal 
black hole entropy can be rewritten as,\footnote{We have changed the overall
sign of this expression so that the result matches with that of \cite{0009234}.}
\be\label{esextremal}
S_{extremal}=-i\, \pi \, Q_I \, \gamma_N^I - 2\, \pi\, i\, F(\{i\gamma_N^I\}, -64)
+i\, \pi \, Q_I \, \gamma_S^I + 2\, \pi\, i \, \bar F(\{-i\gamma_S^I\}, -64)\, ,
\ee
where $F(\{i\gamma_N^I\},-64)$ means in the argument of $F$, $X^I$ is replaced by
$i\gamma_N^I$ and ${\hat A}$ is replaced by $-64$
and similarly for $\bar F(\{-i\gamma_S^I\}, -64)$. 
This was also conjectured in \cite{2402.03297} 
to be the formula for the index.

We shall now bring this expression to a form that will be more useful for us.
First note that due to the
new attractor equation \refb{e242new} and the homogeneity of $F$, we have,
\be
F_I(\{i\gamma_N^I\},-64) = i\gamma_{NI}, \qquad \bar F_I(\{-i\gamma_S^I\},-64)=-i\gamma_{SI}\, .
\ee
We also have, as a consequence of the homogeneity condition,
\ben
2 F(\{i\gamma_N^I\},-64)&=& 
i\gamma_N^I F_I(\{i\gamma_N^I\},-64) -128\, F_{\hat A}(\{i\gamma_N^I\},-64)
\nonumber \\ &=&
-\gamma_N^I \gamma_{NI}  -128\, F_{\hat A}(\{i\gamma_N^I\},-64), 
\nonumber \\
2 \bar F(\{-i\gamma_S^I\},-64)&=&
-i\gamma_S^I \bar F_I(\{-i\gamma_S^I\},-64) -128\, \bar
F_{\hat A}(\{-i\gamma_S^I\},-64)
\nonumber \\ &=& 
- \gamma_S^I
\gamma_{SI}-128\, \bar F_{\hat A}(\{-i\gamma_S^I\},-64)\, . 
\een
Using this, and $Q_I=\gamma_{NI}+\gamma_{SI}$ we can rewrite \refb{esextremal} as,
\ben\label{esextremal2}
S_{extremal}&=&-i\pi \, (\gamma_{NI}+\gamma_{SI}) \, \gamma_N^I 
+ \pi\, i\, (\gamma_N^I \gamma_{NI} +128\, F_{\hat A}(\{i\gamma_N^I\},-64))\nonumber \\ &&
+i\pi \, (\gamma_{NI}+\gamma_{SI})  \, \gamma_S^I - \pi\, i \,(\gamma_S^I \gamma_{SI} 
+128\, \bar F_{\hat A}(\{-i\gamma_S^I\},-64)) \nonumber \\
&=& -i\, \pi \langle \gamma_N,\gamma_S\rangle
+128 \pi i \left[F_{\hat A}(\{i\gamma_N^I\}, -64) - \bar 
F_{\hat A}(\{-i\gamma_S^I\},-64)\right] \, .
\een
Our goal will be to prove that this formula for the spherically
symmetric extremal black hole entropy
agrees with the formula 
\refb{eindex} for the index.

Since various formul\ae\ involve the combinations
$\bar h X^I$ and $h\bar X^I$, it will be useful to define\cite{0009234}:
\be \label{eredef1}
Y^I \equiv \bar h e^{-g} X^I, \qquad \U = \bar h^2 \, e^{-2g}\, {\hat A}\, .
\ee
\refb{ehomx} then gives,
\be \label{eredef2}
F(\Y,\U) = \bar h^2 \, e^{-2g}\,  F(X,{\hat A})\, .
\ee
We shall also define: 
\ben\label{eredef3}
&& F'_I={\p F(\Y,\U)\over \p \Y^I} = \bar h\, e^{-g}\, F_I, 
\qquad F'_\U = {\p F(\Y,\U)\over \p \U}
=F_{\hat A}, \nonumber \\
&& F'_{\U I} = {\p^2 F(\Y,\U)\over \p \U \p \Y^I} 
= \bar h^{-1} \, e^{g}\,  F_{{\hat A}I}, \qquad 
F'_{\U\U}= {\p^2 F(\Y,\U)\over \p \U^2} =\bar h^{-2} e^{2g}\, F_{{\hat A}{\hat A}}\, .
\een

\sectiono{Computation of the Index}

In this section we shall compute the Wald entropy and the angular momentum
associated with the solution described in the last section and then compute the
index via \refb{eindex}.

\subsection{Simplification of the solution on the horizon} \label{ssimpli}

We begin by noting some simplification in the expressions for $T_m$ for the
solutions considered here.
We first note that on the horizon,
\be\label{e3.1xx}
\hat\nabla_1 g = \hat E_1^{~ r}\,  \p_r g =0, \qquad \hat\nabla_3 g = \hat E_3^{~\phi}\p_\phi g=0\, ,
\ee
since $ \hat E_1^{~r}$ vanishes on the horizon and $\p_\phi g=0$ due to axial
symmetry. In arriving at this result we have used the fact that $\p_r g$ is finite at the
horizon -- this has been discussed in appendix \ref{sa}.
Furthermore
we have, from \refb{e511x},
\be\label{ekrel}
K_1 = e^{2g}\, \hat E_2^\theta \hat E_3^{~\phi} \, (\p_\theta\sigma_\phi
-\p_\phi\sigma_\theta)=0, \qquad K_3 = e^{2g}\, \hat E_1^r \hat E_2^{~\theta} \, 
(\p_r\sigma_\theta
-\p_\theta\sigma_r)=0\, ,
\ee
since $\sigma_\beta$ is non-zero only for $\beta=\phi$, and
$\p_\theta\sigma_\phi=\p_\theta \omega_\phi=0$
on the horizon due to \refb{e125x}. Actually since $\hat E_3^{~\phi}$ diverges
as $(r-b)^{-1/2}$ as we approach the horizon, we need a slightly stronger result,
proved in appendix \ref{sa}, that $\p_\theta \omega_\phi$ vanishes as $(r-b)$ as
we approach the horizon.
\refb{e3.1xx} and \refb{ekrel} give, from \refb{e511x},
\ben \label{etrel}
&& T_1 = \bar T_1=0 \, , \quad T_2 = \hat\nabla_2 g -  {i\over 2} K_2, \quad K_2 =  
-e^{2g}\, \hat E_1^{~r} \hat E_3^{~\phi}
\p_r\, \omega_\phi =- {e^{2g}\over b\sin^2\theta} \p_r\omega_\phi
\, , \nonumber \\ && T_3=\bar T_3 = 0\, , 
\quad \bar T_2 = 
\hat\nabla_2 g +  {i\over 2}\, K_2
\, , \qquad \qquad \hbox{for $r=b$.}
\een

For future use, we need to evaluate $\omega_\phi$, $\p_r\omega_\phi$,
$e^{4g}$ and $\p_\theta e^{2g}$ near
the poles. First, by examining  \refb{e233xy} near the poles,  and using
\refb{e242}, \refb{e226x} and the $\chi=-2$ gauge condition,
we get,
\be
-\left\langle \hh + {\gamma_N\over r_N} + {\gamma_S\over r_S}, \, 
d\left(\hh + {\gamma_N\over r_N} + {\gamma_S\over r_S}\right) \right\rangle
+ \star \,  d\sigma = 0\, ,
\ee
where $\star$ denotes three dimensional Hodge dual. This gives,
using  \refb{ea2y}, \refb{ehhgam},
\be
\sigma = \omega_\phi d\phi\, ,
\ee
where,
\be\label{e333x}
\omega_\phi =\langle \gamma_N, \gamma_S\rangle \,
{b\sin^2\theta\over r^2-b^2\cos^2\theta} + \langle \hh, \gamma_N\rangle
\left( {r\cos\theta - b\over r - b\cos\theta}-{r\cos\theta + b\over r + b\cos\theta}
\right)\, .
\ee
This gives,
\be\label{e3.7xx}
\omega_H\equiv \omega_\phi|_{r=b} = b^{-1}\, \langle \gamma_N, \gamma_S\rangle -2\,
\langle \hh, \gamma_N\rangle\, ,
\ee
and 
\be\label{e3.8xx}
\p_r\omega_\phi|_{r=b}=-2\, b^{-2} \, \csc^2\theta\, \langle \gamma_N, \gamma_S\rangle
+ 2\, b^{-1} \, (1+\cos^2\theta) \, \csc^2\theta\, \langle \hh, \gamma_N\rangle\, .
\ee
{\it A priori} \refb{e3.7xx} and \refb{e3.8xx} are valid only near the poles. 
But $\theta$
independence of $\omega_\phi$ on the horizon allows us to identify \refb{e3.7xx} as the
value of $\omega_\phi$ anywhere on the horizon.
\refb{e3.8xx}  shows that,
\ben\label{e342x}
\p_r\omega_\phi|_{r=b} &\simeq& 
- 2 b^{-1}\theta^{-2} \omega_H \ \hbox{for $\theta\simeq 0$} 
\nonumber \\ 
&\simeq&
- 2 b^{-1}(\pi-\theta)^{-2} \omega_H \ \hbox{for $\theta\simeq \pi$}
\, .
\een

Next we need to evaluate $\p_\theta g$ and $e^{2g}$ near the poles. Using the fact
that $e^{-2g}$ has a single pole in $r_N$ near the north pole, we get, for small
$r_N$,
\be\label{esmallrn}
e^{-2g} = {C_N\over r_N} = {C_N\over r-b\cos\theta}\, ,
\ee
where $C_N$ is a constant.
On the horizon $r=b$, this gives,
\be\label{epthetag}
\p_\theta g =  {\sin\theta\over 2(1-\cos\theta)}\, ,
\ee
and hence,
\be
\hat\nabla_2 g = \hat E_2^{~\theta} \p_\theta g = {1\over 2b(1-\cos\theta)}
\simeq {1\over b\, \theta^2}\, .
\ee
On the other hand we have,  from \refb{etrel}, \refb{e342x} and \refb{esmallrn}:
\be 
K_2 = -{e^{2g}\over b\sin^2\theta} \p_r\omega_\phi
\simeq b^{-1} C_N^{-1} \theta^{-2} \omega_H \qquad
\hbox{for $r=b$, $\theta\simeq 0$}\, .
\ee
Therefore, on the horizon near the north pole, we have,
\be
T_2 = \hat\nabla_2 g - {i\over 2} K_2
\simeq b^{-1}\theta^{-2} \left(1-{i\omega_H\over 2C_N}\right),
\qquad \bar T_2 \simeq b^{-1}\theta^{-2} \left(1+{i\omega_H\over 2C_N}\right)\, .
\ee
We now get,
\ben
 && \U = \bar h^2\,  e^{-2g}\, T_{ab} \,   T^{ab} = -64\, T_2^2 = -64\, b^{-2} \, \theta^{-4} \left(1-{i\omega_H
\over 2C_N}\right)^2, \nonumber \\
 && \bar \U  = -64\, b^{-2} \, \theta^{-4} \left(1+{i\omega_H
\over 2C_N}\right)^2\, .
\een 
Comparing the expression for $\bar\U$ 
with \refb{e242new}, using $\bar \U=h^2 e^{-2g} \, 
\bar {\hat A}$ and $r_N=b(1-\cos\theta)
= b\theta^2/2$ on the horizon near the north pole, we get,
\be\label{eomnorm}
\left(1+{i\omega_H\over 2C_N}\right)^2 = 0\, .
\ee
This gives,
\be\label{e317x}
C_N = -i\omega_H/2\, .
\ee
With this result we get, near $r_N=0$,
\be
\U  \simeq -256\, b^{-2} \, \theta^{-4} \simeq - 64\, r_N^{-2}\, ,
\ee
in agreement with \refb{e242new}.

Analysis near the south pole is related to the one described above by a 
$\theta\to \pi-\theta$ transformation. At the south pole, the new attractor equations
require the coefficient of the $1/r_S^2$ term in $\U$ to vanish.
We get,
\be\label{e319x}
C_S =-  i\omega_H/2\, .
\ee
There are two extra minus  signs that cancel.
One originates from the expression for $\hat\nabla_2 g$
which now goes as $-b^{-1}(\pi-\theta)^{-2}$ near the south pole. 
The other one comes from the fact that now the coefficient of $r_S^{-2}$ in
$\U$ has to vanish instead of in $\bar\U$.
As a result, we get,
\be\label{eomnorm_S}
\left(1+{i\omega_H\over 2C_S}\right)^2 = 0\, ,
\ee
which gives \eqref{e319x}. Thus \refb{esmallrn} and \refb{epthetag} are replaced by:
\be \label{esouthg}
e^{-2g} =  -{i\omega_H\over 2(r+b\cos\theta)}\, , \qquad
\p_\theta g|_{r=b} =  -{\sin\theta\over 2(1+\cos\theta)}\, , \qquad \hbox{as $r_S\to 0$}.
\ee

Eqs.~\refb{esmallrn} and 
\refb{e317x} show that in order to get a positive value of $e^{-2g}$, we need to
have $-i\omega_H$ to be positive. Eqs.~\refb{eombeta} and \refb{eAng} 
now give
\be
\beta= {2\pi i\over \Omega_H }=-2\pi i\omega_H >0 \, ,
\ee
as desired.

\subsection{Wald entropy} \label{swald}

We now turn to the computation of the Wald entropy.
For reasons that will become clear later, it will be convenient to carry out a 
field redefinition
\be
D' = D -{1\over 6}\, R
\ee
and use $D'$ as the independent field. Since all physical quantities are invariant
under field redefinition, this step is allowed but not necessary, --
we could carry out our
analysis using the original variable $D$ as well. This will be done in section \ref{s4}.
In terms of the field $D'$ we have,
\ben\label{e315x}
8\pi\LL &=& i F_I \bar X^I \, D'
+{1\over 2} \, i\, \wh F^{-ab} \, F_{{\hat A}I} \, \left(F^{-I}_{ab} -{1\over 4}
\bar X^I \, T_{ab}\right) +{1\over 2} i \, F_{\hat A}\, \wh C
+{i\over 4}F_{{\hat A}{\hat A}} \wh F^{-ab} \, \wh F^{-}_{ab}+ \rm{c.c.}\nonumber \\
&+& \chi\left({1\over 4}R +{1\over 2} D'\right) + \cdots\, ,
\een
\be\label{e115x}
f_\mu{}^a  = {1\over 2} R_\mu^{~a} - {1\over 4} \left(D'+{1\over 2}R\right) e_\mu{}^a+\cdots,
\ee
\be\label{e116x}
\mathcal{R}(M)_{ab}{}^{cd}=R_{ab}{}^{cd}-2R_{[a}^{[c}\delta^{d]}_{b]}+{R\over 2}\delta^c_{[a}\delta^d_{b]}
+D'\delta^c_{[a}\delta^d_{b]}+ \cdots\, .
\ee
Computation of the Wald entropy will require taking derivative of the action with respect
to the Riemann tensor. We shall take the derivative at fixed $D'$. Since the Wald
entropy is invariant under a field redefinition, the final result should be independent of
whether we take the derivative keeping $D$ fixed or $D'$ fixed. 
In the $\chi=-2$ gauge 
the coefficient of the $\int d^4x\sqrt{-\det g} R$ term  in the action now becomes
$-1/(16\pi)$. This corresponds to setting $G_N=1$.

For computing the Wald entropy\cite{9307038,9312023}, we need the 
binormal to the horizon at $r=b$, $t= \mathrm{constant}$. This has the form of an anti-symmetric
tensor $\ve_{\mu\nu}$ whose only non-zero components are $\ve_{rt}$ and $\ve_{tr}$
and which is normalized as $\ve_{\mu\nu}\ve^{\mu\nu}=-2$.
It will be convenient  to express this using tangent space indices.   
This has the form
\be
\ve_{ab} = E_a^{~t} E_b^{~r} \ve_{tr} - (a\leftrightarrow b)\, , \quad \quad \quad \ve_{tr} = \frac{\sqrt{-\det g}}{\sqrt{\det g_H}},
\ee
where $g_H$ is the induced metric on the horizon. 
Examining the form of the inverse vielbeins given in \refb{einvfour} 
we see that the only non-zero
components of $\ve_{ab}$ are $\ve_{13}$ and $\ve_{31}$. We take 
\be \label{binormal}
\varepsilon_{13}=\varepsilon^{13}=i,\, \qquad \varepsilon_{31}=\varepsilon^{31}=-i\, .
\ee
They satisfy the normalization condition
$\ve^{ab}\ve_{ab}=-2$. With this normalization, the Wald entropy is  given by,
\be\label{eww10}
S_{wald}={\pi\over 2} \, \ve_{ab} \ve_{cd} 
\int_H {\p \LL \over \p R_{abcd}}\,  d\theta \, d\phi \, {\sqrt{\det g_H}}\, ,
\ee
 where the derivative with respect to the Riemann tensor is defined with the
 normalization
 \be\label{e210x}
{\p R_{a b c d } \over \p R_{pqrs}} 
= \delta^p_a  \, \delta^q_b  \, \delta^r_c  \delta^s_d 
- \delta^q_a  \, \delta^p_b  \, \delta^r_c  \delta^s_d 
-\delta^p_a  \, \delta^q_b  \, \delta^s_c  \delta^r_d 
+\delta^q_a  \, \delta^p_b  \, \delta^s_c  \delta^r_d \, .
\ee
It can be easily checked that for the Einstein-Hilbert action $-(16\pi G_N)^{-1}
\int \sqrt{-\det g} \, R \, d^4x$,  \refb{eww10} 
produces the standard Bekenstein-Hawking entropy.
For the metric we are considering, \refb{eww10} reduces to:
\be\label{eww11}
S_{wald}= {i\over 4} \int_H {\p (8\pi \LL) \over \p R_{3131}}
\, d\theta \, d\phi \, \omega_H\, b\, \sin\theta\, .
 \ee
The choice of  sign in this expression requires an explanation since it depends
on the sign of $\sqrt{\det g_H}$.
Using \refb{eombeta}, \refb{eAng} and the positivity of $\beta$ we see that
$-i\omega_H$ should be positive.  Therefore $\sqrt{\det g_H}$ in
\refb{eww10} should be
proportional to $-i\omega_H$ with a positive constant of proportionality. 
The $(\eps_{13})^2$ factor in \refb{eww10} gives an additional minus sign. We also note that the choice 
\refb{binormal} is consistent with $\ve_{tr} > 0$. 

We write \refb{eww11} as
\be \label{ewaldin}
S_{wald}={i\over 4}\, \omega_H \, b\,
\int_H \II\,  \sin\theta\, d\theta\,
d\phi\, ,
\ee
where we use \refb{e315x} to write, in the tangent space notation
\ben \label{e530x}
\II &\equiv& 8\pi \, {\p\LL\over \p R_{3131}}\nonumber \\
&=& {1\over 4} \chi {\p R \over \p {R_{3131}} }
+\left[{1\over 2} i \, F_{\hat A}\, {\p \wh C\over \p R_{3131}} +{\rm c.c.}\right]
  + \left[{1\over 2} \, i\, {\partial \wh F^{-ab}\over \p R_{3131}} \, F_{{\hat A}I} 
  \left(F^{-I}_{ab} -{1\over 4}
\bar X^I \, T_{ab}\right) + \hbox{c.c.}
\right]
\nonumber \\
&+&
\left[ {i\over 2} F_{{\hat A}{\hat A}} {\p \wh F^{-ab}\over \p R_{3131}} \wh F^-_{ab}  +\hbox{c.c.}\right]
\nonumber \\
&=& \II_0+\II_1+\II_2+\II_3\, ,
\een
where
\ben\label{e213x}
\II_0 &=&  {1\over 4}\chi {\p R \over \p {R_{3131}} } \, , \nonumber \\ 
\II_1 &=&
\left[ {64\, i} \, F'_\U \, {\p \RR(M)^{-cdab}\over \p R_{3131}}\, \RR(M)^-_{cdab}
+8 i\, F'_\U\,  {\p f_a^{~c}\over \p R_{3131}} \, T^{ab}\,  \bar T_{cb}
+\hbox{c.c.}\right],\nonumber \\ 
\II_2 &=&
\left[ -8\, i\, {\p \RR(M)^{cdab}\over \p R_{3131}} \bar h\, T_{cd}
\, e^{-g}\, F'_{\U I} \left(F^{-I}_{ab} -{1\over 4}
e^g \, \bar \Y^I \, \bar h\, T_{ab}\right)  + \hbox{c.c.}\right],\nonumber \\
\II_3 &=& \left[ 128\, i\, e^{-2g}\, 
F'_{\U\U} \, {\p \RR(M)^{cdab}\over \p R_{3131}} \, \bar h\, T_{cd} 
\, \RR(M)^{c'd'}_{~~~ab} \, \bar h\, T_{c'd'} +\hbox{c.c.}
\right] .
\een
In arriving at \refb{e213x} we have used \refb{eredef1}-\refb{eredef3}. In the rest of the section, we shall evaluate all quantities
on the horizon unless mentioned otherwise.

Evaluation of $\II_0$ is straightforward. Using \refb{e210x} we get,
\be\label{ei0x}
\II_0 = {1\over 2}\, \chi = -1\, ,
\ee
where we used the gauge condition $\chi=-2$.

Next we turn to $\II_1$.
Using \refb{e210x} and the definitions of $f_{ac}$, $\RR(M)_{abcd}$ given in
\refb{e115x}, \refb{e116x}, we get,
\be\label{efderivative}
{\p f_{a c }\over \p R_{3131}} 
= {1\over2} \left[\delta^3_a  \, \delta^3_c 
+  \delta^1_a  \, \delta^1_c \right]
-{1\over 4}
\, \eta_{a c }
\, ,
\ee
\ben \label{e535x}
{\p \RR(M)_{a b c d }\over \p R_{3131}}
&=& \delta^3_a  \, \delta^1_b  \, \delta^3_c  \delta^1_d 
- \delta^1_a  \, \delta^3_b  \, \delta^3_c  \delta^1_d 
-\delta^3_a  \, \delta^1_b  \, \delta^1_c  \delta^3_d 
+\delta^1_a  \, \delta^3_b  \, \delta^1_c  \delta^3_d \nonumber \\
&-&
\Bigg[\left\{{1\over 2} \left( \delta^3_a  \, \delta^3_c 
+  \delta^1_a  \, \delta^1_c \right) \eta_{b d} 
-{1\over 4} 
\, \eta_{a c } \eta_{b d}\right\} \nonumber \\ &&
- (a \leftrightarrow b ) - (c \leftrightarrow d)
+(a \leftrightarrow b , c \leftrightarrow d)\Bigg]\, .
\een
Using \refb{efderivative}, \refb{e535x}, 
\refb{e509}-\refb{e511}, \refb{e512}, the anti-symmetry of
$\RR(M)^-_{abcd}$ in the first two indices and the last two indices, and the vanishing of $T_1$ and $T_3$ on the horizon from
\refb{etrel}, we now get,
\ben\label{e539}
{\p \RR(M)^-_{a b c d}\over \p R_{3131}} \RR(M)^{-a b c d}
&=& 2 \RR(M)^-_{31 31} - 2 \RR(M)^-_{0202} \nonumber \\
&=& \RR(M)_{31 31} - \RR(M)_{0202} - i (\RR(M)_{0231}+\RR(M)_{3102})\nonumber \\
&=& e^{2g} \left(2\hat\nabla_2 T_2  + 2 T_2^2\right)\, ,
\een
and
\be\label{e541}
{\p f_{ab}  \over \p R_{3131}} \, T^{ac}\,  \bar T^b_{~c}
= 16\, e^{2g}\, \bar T_2 T_2\, .
\ee
Using \refb{e213x}, \refb{e539} and \refb{e541} we now get:
\be\label{eI1}
\II_1 = 64\, i \, F'_\U \, e^{2g} \, \left(2\hat\nabla_2 T_2  + 2 T_2^2
+ 2\, \bar T_2 T_2 \right) + {\rm c.c.}\, .
\ee
Using \refb{etrel} this can be simplified to:
\be
\II_{1} = 128 \, i\, F'_\U \,  \hat E_2^{~\theta}  \p_\theta \left( e^{2g} T_2
\right)  +{\rm c.c.} \, .
\ee
This can be written as,
\ben
\II_{1} &=& 128 \, i\,  \hat E_2^{~\theta}  \p_\theta \left( e^{2g} T_2 \, F'_\U \, 
\right) - 128 \, i\,   e^{2g} T_2 \, F'_{\U I}   \hat\nabla_2 \Y^I\, 
-  128 \, i\, e^{2g}\, T_2 \, F'_{\U\U}  \hat\nabla_2 \U +{\rm c.c.} \nonumber \\
&=& 128 \, i\,  \hat E_2^{~\theta}  \p_\theta \left( e^{2g} T_2 \, F'_\U \, 
\right) - 128 \, i\,   e^{2g} T_2 \, F'_{\U I}   \hat\nabla_2 \Y^I\, 
 -  \, 256 \, i\,  e^{2g} \U \, F'_{\U\U}  \hat\nabla_2 T_2 
+{\rm c.c.}, \nonumber \\
\een
where in the last step we have used,
\be \label{e563}
\U = \bar h^2\,  e^{-2g}\, T_{ab} \,  T^{ab}
= -64\, T_2^2\, .
\ee

Next we  consider the term $\II_2$ given in \refb{e213x}. Using \refb{e535x}, 
\refb{e509}-\refb{e511}, \refb{e135x} 
and \refb{etrel} this can be
reduced to,
\ben\label{e559x}
\II_2 &=& 128\, i\, F'_{\U I}\, e^{g}\, T_2\, 
\left(\hat \nabla_2 (e^g \Y^I)  -{1\over 2} \, i\, K_2\, e^g\, \Y^I\right)
+ {\rm c.c.} \nonumber \\
&=&128\, i\, F'_{\U I}\, e^{2g}\, T_2\, 
\hat \nabla_2 \Y^I 
- 256 \, i\,  \, \U\, F'_{\U\U}\, e^{2g}\, T_2^2 + {\rm c.c.} \, ,
\een
where in the last step we have used the homogeneity property,
\be
F'_{\U I} \Y^I + 2\, F'_{\U\U} \, \U=0\, .
\ee

Finally we turn to the term $\II_3$ given in \refb{e213x}.
Using \refb{e535x}, \refb{e509}-\refb{e511} and \refb{e512} we can express this as 
\be\label{eI3}
\II_3 = 256\, i\, e^{2g}\, \U\, F'_{\U\U}  \,  (\hat\nabla_2 T_2 + T_2^2) +{\rm c.c.}\, ,
\ee
where we have used \refb{e563} again.

Adding the contributions from $\II_0$,
$\II_1$, $\II_2$ and $\II_3$, and using \refb{etrel} and
\refb{e563}, we get,
\ben \label{esumI}
&& \II_0+ \II_1+\II_2+\II_3 = -1+ \left[128 \, i\,  \hat E_2^{~\theta}  
\p_\theta \left( e^{2g} T_2 \, F'_\U \, 
\right) + {\rm c.c.} \right]  \nonumber \\
&=& -1+ \left[ 128 \, i\,  {1\over b\sin\theta}\,  
\p_\theta \left( e^{2g} T_2 \, F_{\hat A} 
\right)- 128 \, i\,  {1\over b\sin\theta}\,  
\p_\theta \left( e^{2g} \bar T_2 \, \bar F_{\hat A} 
\right) \right] \, ,
\een
where we have used the relations $F'_\U=F_{\hat A}$, $\bar F'_\U=\bar F_{\hat A}$.
Using \refb{ewaldin} and \refb{etrel} we now get,
\ben \label{ewaldfin}
S_{wald}&=&- i\pi \, \omega_H\, b   - {32 \,\pi} \, \omega_H\, \Bigg[
 \Bigg[F_{\hat A}\,\left({1\over b\sin\theta}
\p_\theta e^{2g}
+ i\, {e^{4g}\over b \sin^2\theta} \p_r\omega_\phi\right)
 \Bigg]_0^\pi  \nonumber \\ && \hskip 1in
 - \Bigg[\bar F_{\hat A}\,\left({1\over b\sin\theta}
\p_\theta e^{2g}
- i\, {e^{4g}\over b \sin^2\theta} \p_r\omega_\phi\right)
 \Bigg]_0^\pi
 \Bigg]_{r=b}\, .
\een
This can be evaluated using the behaviour of $\omega_H$, $\p_r\omega_\phi$,
$e^{2g}$ and $\p_r g$ near the poles, as derived in section \ref{ssimpli}.
The result is:
\be \label{swald2}
S_{wald}=- i\pi \,  ( \langle \gamma_N, \gamma_S\rangle -2\, b\, 
\langle \hh, \gamma_N\rangle)  + \left[
128\, \pi\,  i \,   F_{\hat A} |_N
- 128\, \pi\,  i \,  \bar F_{\hat A} |_S\right]\, .
\ee

\subsection{Angular momentum and index}

We also need to compute the angular momentum $J$. For this we examine \refb{e233x}
for large $r$. In this limit $g\to 0$, $\p g \to r^{-2}$, 
the metric approaches flat metric, and the right hand
side, containing two extra derivatives, is suppressed compared to the left hand side.
Therefore \refb{e333x} is valid even in the asymptotic region, and we get, taking the
large $r$ limit of \refb{e333x},
\be
\omega_\phi\simeq - {2\, b\, \sin^2\theta\over r}\, \langle \hh, \gamma_N\rangle .
\ee
It now follows from the form of the metric given in \refb{emetric} with 
$\sigma=\omega_\phi\, d\phi$ that the angular
momentum carried by the black hole is given by,
\be\label{eJ}
J = {r\, \omega_\phi \over 2\, G_N \sin^2\theta} = - b\, \langle \hh, \gamma_N\rangle \, ,
\ee
using the fact that for the $\chi=-2$ gauge choice, $G_N=1$.

Using \refb{swald2} and \refb{eJ} we get the expression for the index:
\be
S_{wald} + 2\pi i J = - i\pi \,  \langle \gamma_N, \gamma_S\rangle  + \left[
128\, \pi\,  i \,   F_{\hat A} |_N
- 128\, \pi\,  i \,  \bar F_{\hat A}|_S\right]\, .
\ee
This agrees with \refb{esextremal2}, establishing the equality of the index and the
entropy of these black holes.

Note that this analysis assumes the existence of a solution to \refb{e232x}, 
\refb{e233x} that 
interpolates between the asymptotically flat space-time and the near horizon 
configuration described in section \ref{ssimpli}. 
We believe that such a solution should
exist even though we have not proved it.

\sectiono{An alternate derivation of the Wald entropy} \label{s4}

In section \ref{swald} we used \refb{e315x} to compute the Wald entropy, treating $D'$ as
the independent variable. In this section we shall use the original form of the
action, treating $D$ as the independent variable, to compute the Wald entropy.
Since the two actions are related by field redefinition, we expect the final results to
agree. Nevertheless the intermediate steps are different, and this analysis can be
regarded as a confirmation that Wald entropy is invariant under field redefinition.

In this approach the integrand $\II$ in the Wald entropy expression \refb{ewaldin}
takes the form:
\ben
\II &\equiv& 8\pi \, {\p\LL\over \p R_{3131}}\nonumber \\
&=& {1 \over 6} (\chi - e^{-K}){\p R \over \p  {R_{3131}} }\nonumber\\
&+&\left[{1\over 2} \, i\, {\p \wh F^{-ef}\over \p R_{3131}} \, F_{\hat{A}I} \left(F^{-I}_{ef} -{1\over 4}
\bar X^I \, T_{ef}\right) +{1\over 2} i \, F_{\hat{A}}\, {\p \wh C\over \p R_{3131}} + \hbox{c.c.}
\right]
\nonumber \\
&+&
\left[ {i\over 2} F_{\hat{A}\hat{A}} {\p \wh F^{-ef}\over \p R_{3131}} \wh F^-_{ef}  +\hbox{c.c.}\right]\, ,
\een
where,
\be
e^{-K} = -i (X^I \bar F_I-\bar X^I F_I)\, .
\ee

We shall evaluate various terms using the various relations given in 
sections \ref{s2} and \ref{swald}. The analogs of \refb{efderivative} and \refb{e535x} now take the form,
\be\label{efderivative_2}
{\p f_{a c }\over \p R_{3131}} 
= {1\over2} \left[\delta^3_a  \, \delta^3_c 
+  \delta^1_a  \, \delta^1_c \right]
-{1\over 6}
\, \eta_{a c }
\, ,
\ee
\ben \label{e535x_2}
{\p \RR(M)_{a b c d }\over \p R_{3131}}
&=& \delta^3_a  \, \delta^1_b  \, \delta^3_c  \delta^1_d 
- \delta^1_a  \, \delta^3_b  \, \delta^3_c  \delta^1_d 
-\delta^3_a  \, \delta^1_b  \, \delta^1_c  \delta^3_d 
+\delta^1_a  \, \delta^3_b  \, \delta^1_c  \delta^3_d \nonumber \\
&-&
\Bigg[\left\{{1\over 2} \left( \delta^3_a  \, \delta^3_c 
+  \delta^1_a  \, \delta^1_c \right) \eta_{b d} 
-{1\over 6} 
\, \eta_{a c } \eta_{b d}\right\} \nonumber \\ &&
- (a \leftrightarrow b ) - (c \leftrightarrow d)
+(a \leftrightarrow b , c \leftrightarrow d)\Bigg]\, .
\een
Let us first consider,
\be
\II_1'={i\over 2}F_{\hat{A}}{\p \wh C\over \p R_{3131}}+\hbox{c.c.} \, .
\ee
 The contribution from \refb{e541} is unchanged since \refb{efderivative_2} only differs in the last term from \refb{efderivative}, and the difference vanishes since $T_{ab}\bar{T}^{ab}=0$. The analog of \refb{e539} reads,
\ben
{\p \RR(M)^-_{a b c d}\over \p R_{3131}} \RR(M)^{-a b c d}
&=&4 \RR(M)^-_{31 31} - {\widetilde{D}} \nonumber \\
&=& e^{2g} \left(2\hat\nabla_2 T_2  + 2 T_2^2\right)- {\widetilde{D}}\, ,
\een	
where in going from first line to the second we have used \refb{e512}. In writing the first line above, we have performed the contractions arising from the second and third line of \refb{e535x_2} using a relation that defines $\widetilde{D}$,
\be\label{eRM-trace}
\RR(M)^-_{abcd}\eta^{bd}+(a \leftrightarrow c)= {3\over 2} \, \widetilde D\, \eta_{ac}.
\ee
$\widetilde{D}$ can be evaluated from \refb{e512} to be,
\ben\label{eDexp}
\widetilde{D}&=&{2\over 3}e^{2g}\left(\hat{\nabla}^pT_p-T_p^2\right) \\
&=& {2\over 3}e^{2g}\left(\hat{\nabla}^p \bar{T}_p-\bar{T}_p^2\right) \\
&=& {2\over 3}e^{2g} \left[\hat \nabla_p^2 g - (\hat \nabla_p g)^2 + \frac{1}{4} e^{4g} (R(\sigma)_p)^2\right],\label{eDexp3}
\een
where the second equality  can be seen using \refb{e511x} and $\hat \nabla^pR(\sigma)_p=0$. It follows from \cite{0009234} (eq.~(5.31) or eq.~(B.5)) that
$\widetilde D$ is the same as the auxiliary field $D$, but we shall not need to
use this information. 

Thus $\II_1'$ evaluates to, 
\ben 
\II_1^\prime&=&-64 \, i \, F^\prime_\Upsilon \, \widetilde{D} + 128 \, i \, F^\prime_\Upsilon \, e^{2g} \, \left(\hat{\nabla}_2T_2+T_2^2+T_2\bar{T}_2\right)+\hbox{c.c.}\nonumber\\
&=&-64\, i \, F^\prime_\Upsilon \, \widetilde{D}+\hbox{c.c.}+\II_1,
\een
where in the second line we have extracted the contribution $\II_1$ that matches with \refb{eI1}.

Next we consider the term proportional to $(\chi-e^{-K})$ in $\II$:
\be
 \II_0^\prime={1 \over 6} \left(\chi - e^{-K}\right){\p R \over \p {R_{3131}} }\nonumber\\
=-{1\over 3}(2+e^{-K})\nonumber\\
=-1-\left({{e^{-K}-1}\over 3}\right),
\ee
where we have used $\chi=-2$. The first term 
on the right hand side 
is the integrand for the area term and agrees with $\II_0$ given in \refb{ei0x}. 
We will denote the second term as,
\ben
 \II_0''=-\left({{e^{-K}-1}\over 3}\right).
\een
From \refb{e232x}, we have,
\ben
e^{-K}-1 &=& -128\, i\, e^{3g}\, \hat\nabla^p [e^{-g} \hat\nabla_p g (F_{\hat A}-\bar 
F_{\hat A})]
- 32\, i\, e^{6g} (R(\sigma)_p)^2 (F_{\hat A}-\bar F_{\hat A}) \nonumber \\ &&
- 64\, e^{4g}\, R(\sigma)_p \hat\nabla^p (F_{\hat A}+\bar F_{\hat A})\, .
\een 
Combining this with \refb{eDexp3} we get on the horizon,
\ben
e^{-K}-1=-192\, i\, F'_\U \, \widetilde D -128\, i \, e^{2g}\, T_2 \, F'_{\U I}\, \hat\nabla_2 \Y^I -128\, i \, e^{2g}\, T_2  \,
F'_{\U\U}\, \hat\nabla_2\U +{\rm c.c.},
\een
where we have expressed all the terms in terms of $\Y^I,\U$.  This gives,
\ben
\II_0''&=&64\, i \, F'_\U \, \widetilde D+{128\over 3} \, i \, e^{2g} \, T_2 \, F'_{\U I}\,\hat\nabla_2 \Y^I +{128\over 3}\,
i \, e^{2g}\, T_2 
\, F'_{\U\U} \, \hat\nabla_2\U + {\rm c.c.}\, \nonumber\\
&=&64\, i \, F'_\U \, \widetilde D+{\rm c.c.}+{1 \over 3}\left(\II_2+\II_3\right),
\een
where $\II_2$ and $\II_3$ are from \refb{e559x} and  \refb{eI3}.
We see that the terms proportional to $\widetilde D$ in $\II_1'$ and $\II_0''$ cancel.

Let us now compute the term proportional to
$F_{\hat A I}$ in $\II$. We get
\ben
\II_2'&=&{i\over 2} \, {\p \wh F^{-ef}\over \p R_{3131}} \, 
F_{\hat{A}I} \, \left(F^{-I}_{ef} -{1\over 4}
\bar X^I \, T_{ef}\right) +\hbox{c.c.}\nonumber\\
&=& {256i \over 3} \,  F_{\Upsilon I}' \,
e^{2g}  \, T_2 \, \left(T_2\Y^I+\hat\nabla_2\Y^I\right) +\hbox{c.c.} \, \nonumber\\
&=&{2\over 3} \, \II_2,
\een
where $\II_2$ is from \refb{e559x}. Finally,
the $F_{\hat A\hat A}$ term in $\II$ gives
\ben
\II_3'&=& {i\over 2} \, F_{\hat{A}\hat{A}} \, {\p \wh F^{-ef}\over \p R_{3131}} 
\, \wh F^-_{ef}  +\hbox{c.c.} \nonumber\\
&=& -
{{2\times (128)^2i}\over 3} \, e^{2g} \, F^\prime_{\Upsilon\Upsilon} \, T_2^2 \,
\left(\hat\nabla_2T_2 +T_2^2\right)
+ \hbox{c.c.} \nonumber\\
&=&{2\over 3} \, \II_3,
\een
where $\II_3$ is from \refb{eI3}. Adding all the contributions, we get
$
\sum_{i=0}^3\II_i'=\sum_{i=0}^3\II_i.
$
This reproduces \refb{esumI} after using \refb{etrel}.

\bigskip

\noindent{\bf Acknowledgement}: 
We thank S.~Adhikari, S.~Chakraborty, S.~Kashyap, and S.~Sarkar for discussions. 
S.H. is funded by the European Union (ERC, UNIVERSE PLUS, 101118787). 
Views and opinions expressed are however those of the authors only and do 
not necessarily reflect those of the European Union or the European Research 
Council Executive Agency. Neither the European Union nor the granting authority 
can be held responsible for them. A.S. is supported by ICTS-Infosys 
Madhava Chair Professorship and 
the Department of Atomic Energy, Government of India, under project no. RTI4001.
S.P. is supported by 
the Department of Atomic Energy, Government of India, under project no. RTI4001. 
The work of A.V. was partly supported by SERB Core Research Grant CRG/2023/000545 and by the ``Scholar in Residence'' program of IIT Gandhinagar. 

\appendix

\sectiono{Boyer-Lindquist coordinates} \label{sb}

In this appendix we shall review the relation between the 
Boyer-Lindquist coordinates appearing in \refb{embase} and
the Cartesian coordinates of the flat three dimensional base space.
If $x^1,x^2,x^3$ are the Cartesian coordinates then we have
\be
x^1 = \sqrt{r^2-b^2} \, \sin\theta\, \cos\phi, \qquad
x^2 = \sqrt{r^2-b^2} \, \sin\theta\, \sin\phi, \qquad
x^3 = r\,\cos \theta\, .
\ee
Of special interest will be the north and the south poles at the horizon, corresponding
to the points $r=b$ and $\theta=0,\pi$. In the
Cartesian coordinates they correspond to the points $x^1=x^2=0$, $x^3=\pm b$.
We shall denote by $r_N$ and $r_S$ the
distance of any point $(r,\theta,\phi)$ from the north and the south poles respectively.
Then we have,
\ben\label{ea2x}
r_N &=& \sqrt{(x^1)^2 + (x^2)^2 + (x^3-b)^2} = r- b\, \cos\theta, \nonumber \\
r_S &=& \sqrt{(x^1)^2 + (x^2)^2 + (x^3+b)^2} = r + b\, \cos\theta\, .
\een

\sectiono{Estimate of $\p_\theta\omega_\phi$ and $g$ near the horizon} \label{sa}

In this appendix we shall show that $\p_\theta\omega_\phi$ is of order $(r-b)$ 
and $\p_r g$ remains finite as we approach the horizon at $r=b$.
For this we first make a Euclidean continuation of the metric \refb{emetric} 
by setting
$t=-i\tau$ and $\omega_\phi=-i\omega_E$:
\be
ds^2 = e^{2g} (d\tau + \omega_E \, d\phi)^2 + e^{-2g} \left[
{{r^2-b^2\cos^2\theta} \over {r^2-b^2}}dr^2+(r^2-b^2\cos^2\theta)d\theta^2+(r^2-b^2)\sin^2\theta d\phi^2\right]\, .
\ee
Let us define $\omega_0$ to be the value of $\omega_E$ at $r=b$. According to
\cite{2305.08910} this should be independent of $\theta$ for getting a non-singular
metric. Our goal is to determine how fast $\p_\theta\omega_E$ vanishes as
$r$ approaches $b$.

Let us define,
\be
\tilde\phi = \phi + \omega_0^{-1}\tau\, ,
\ee
and express the metric as,
\ben
ds^2 &=& e^{2g} \left(d\tau \left(1- {\omega_E\over \omega_0}\right)
+ \omega_E\, d\tilde \phi\right)^2  + e^{-2g} \left[
{{r^2-b^2\cos^2\theta} \over {r^2-b^2}}dr^2+(r^2-b^2\cos^2\theta)d\theta^2\right]
\nonumber \\ &+& e^{-2g}\, (r^2-b^2)\sin^2\theta\,  (d\tilde \phi- \omega_0^{-1} d\tau)^2
\nonumber \\
&=& d\tilde\phi^2 \left[ e^{2g}\, \omega_E^2 +e^{-2g}(r^2-b^2)\sin^2\theta \right]
+ d\theta^2 \, e^{-2g} (r^2 - b^2\cos^2\theta) + e^{-2g}
{{r^2-b^2\cos^2\theta} \over {r^2-b^2}}dr^2 \nonumber \\ 
&+& d \tau^2\, \left[e^{2g} \left(1- {\omega_E\over \omega_0}\right)^2 
+ e^{-2g}\,  (r^2-b^2)\sin^2\theta\,  \omega_0^{-2}
\right] \nonumber \\
&+& 2\, d\tau d\tilde\phi\left[e^{2g} \omega_E \left(1- {\omega_E\over \omega_0}\right)
- e^{-2g} (r^2-b^2)\sin^2\theta\, \omega_0^{-1}
\right]\, .
\een
This metric appears to have singularity at $r=b$, but following standard procedure we
can try to make the metric manifestly non-singular by making a change of variable,
\be
\rho=\sqrt{r-b}\, .
\ee
Let us suppose that near $\rho=0$, $\omega_E$ and $g$ 
have expansions of the form,
\be
\omega_E = \omega_0(1 + \rho \, f_1(\theta) + \rho^2 f_2(\theta)+\cdots),
\qquad g = g_0(\theta) + g_1(\theta)\, \rho + g_2(\theta) \rho^2 +\cdots\, .
\ee
Then the metric near $\rho=0$ takes the form,
\ben\label{eappmetric}
ds^2 &=& e^{2g_0}\, \omega_0^2\,  (1+2g_1\rho+ 2 \, f_1\rho+\OO(\rho^2)) \, d\tilde\phi^2 
+ e^{-2g_0} \, b^2\sin^2\theta \, (1-2g_1\rho+\OO(\rho^2)) \, d\theta^2 \nonumber \\
&+& 2\, e^{-2g_0} \, b\, (1 -2\, g_1\rho+\OO(\rho^2)) \, \sin^2\theta\, d\rho^2  \nonumber \\
&+& \rho^2 \, d\tau^2\,  \left(e^{2g_0} f_1(\theta)^2 + 2\, b\, e^{-2g_0} \omega_0^{-2}
\, \sin^2\theta +\OO(\rho)\right)\nonumber \\
&-& 2\, d\tau\, d\tilde\phi\, e^{2g_0} \, \omega_0 \, f_1\, \rho \, (1+\OO(\rho))\, .
\een
By examining this metric we see that if $ e^{2g_0} f_1(\theta)$ is proportional to
$\sin\theta$  then at leading order in an expansion in powers
of $\rho$, the metric in the $\rho-\tau$ space is proportional to
$\sin^2\theta(d\rho^2 + c^2 \rho^2\, \tau^2)$ for a constant $c$. We can then
regard this as the polar coordinate system in a two dimensional plane
if $\tau$ is identified as an angular variable with period $2\pi/c$. The corresponding
Cartesian coordinates are related to $\rho,\tau$ via $\rho=\sqrt{x^2+y^2}$ and
$\tau = c^{-1} \tan^{-1}(y/x)$. However we now see that 
the $d\tau d\tilde\phi$ part of the metric \refb{eappmetric} becomes,
\be
-2\, e^{2g_0} \, \omega_0 \, f_1 \, c^{-1} \, 
d\tilde\phi \, {xdy-ydx\over \sqrt{x^2+y^2}} \, .
\ee
This is not analytic at $x=y=0$, as can be seen from the fact that sufficiently
high derivative of $g_{\tilde\phi x}$ and $g_{\tilde\phi y}$ will diverge at 
$x=y=0$. This can be avoided if $f_1$ vanishes. More generally we shall
require the power series expansion of $\omega_E$ to have only even powers
of $\rho$, i.e.\ it should have a Taylor series expansion in powers of $(r-b)$.
Since $\p_\theta\omega_E=0$ at $r=b$, this shows that $\p_\theta\omega_E$ 
is of order $(r-b)$ near the horizon.

Also, in this coordinate system
the $d\tilde \phi^2$ term in the metric \refb{eappmetric} has the form,
\be
d\tilde\phi^2\, e^{2g_0}\, \omega_0^2\,  (1+ 2g_1\, \sqrt{x^2+y^2}+\cdots)\, ,
\ee
where we have used the fact that $f_1$ vanishes. We again see that this is not
analytic at $x=y=0$ since sufficiently high derivative of $g_{\tilde\phi\tilde\phi}$
diverges at $x=y=0$. This can be avoided if $g_1=0$ and more generally $g$
has a Taylor series expansion in $\rho^2$ or equivalently in $(r-b)$. This in turn
shows that $\p_r g$ is finite at $r=b$.

\end{document}